\documentclass[final,5p,times,twocolumn]{elsarticle}

\usepackage{graphics}
\usepackage{amssymb}
\journal{Computer Physics Communications}

\begin{document}

\begin{frontmatter}

%% Title, authors and addresses

%% use the tnoteref command within \title for footnotes;
%% use the tnotetext command for the associated footnote;
%% use the fnref command within \author or \address for footnotes;
%% use the fntext command for the associated footnote;
%% use the corref command within \author for corresponding author footnotes;
%% use the cortext command for the associated footnote;
%% use the ead command for the email address,
%% and the form \ead[url] for the home page:
%%
%% \title{Title\tnoteref{label1}}
%% \tnotetext[label1]{}
%% \author{Name\corref{cor1}\fnref{label2}}
%% \ead{email address}
%% \ead[url]{home page}
%% \fntext[label2]{}
%% \cortext[cor1]{}
%% \address{Address\fnref{label3}}
%% \fntext[label3]{}

\title{Realistic Modeling of Complex Oxide Materials}

%% use optional labels to link authors explicitly to addresses:
%% \author[label1,label2]{<author name>}
%% \address[label1]{<address>}
%% \address[label2]{<address>}

\author{I. V. Solovyev}
\ead{SOLOVYEV.Igor@nims.go.jp}

\address{Computational Materials Science Center,
National Institute for Materials Science, \\
1-2-1 Sengen, Tsukuba, Ibaraki 305-0047, Japan
}

\begin{abstract}
Since electronic and magnetic properties
of many transition-metal oxides
can be efficiently controlled by external factors such as
the temperature, pressure, electric or magnetic field, they are regarded as
promising materials for various applications. From the viewpoint of
electronic structure, these phenomena are frequently related to the behavior of a small
group of states close to the Fermi level.
The basic idea of this project is to construct a low-energy model for the states near the Fermi level on the basis of first-principles density functional theory, and to study this model by modern many-body techniques.
After a brief review of the method, the abilities of this approach will be illustrated on a number of examples,
including multiferroic manganites and spin-orbital-lattice coupled phenomena in $R$VO$_3$
($R$ being the three-valent element).
\end{abstract}

\begin{keyword}
%% keywords here, in the form: keyword \sep keyword
first-principles calculations \sep effective models \sep transition-metal oxides
%% PACS codes here, in the form: \PACS code \sep code
\PACS 71.15.-m \sep 71.10.Fd \sep 75.47.Lx \sep 71.10.-w
%% MSC codes here, in the form: \MSC code \sep code
%% or \MSC[2008] code \sep code (2000 is the default)

\end{keyword}

\end{frontmatter}

%%
%% Start line numbering here if you want
%%
% \linenumbers

%% main text
\section{Basic Idea, Purpose, and Methods of Realistic Modeling}
\label{sec.intro}
\begin{figure}[!h]
\centering
\includegraphics[width=3.0in]{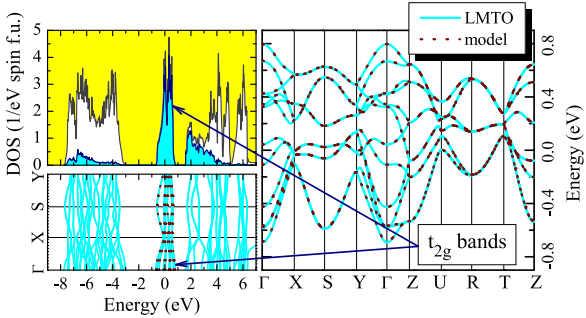}
\caption{(Left panel) Electronic structure of orthorhombic
YVO$_3$ in LDA. The shaded area shows the contributions of the
$3d$ states of V. (Right panel) Enlarged behavior of $t_{2g}$ bands
computed from LMTO basis functions (solid curves)
and downfolded bands (dot-dashed curves).
The corresponding bands in the left panel are shown by arrows.
The
Fermi level is at zero energy.}
\label{fig.YVO3LDA}
\end{figure}

The oxide materials can be rather complex.
Nevertheless, in many cases their
electronic and magnetic properties are
controlled by a small group of states located near the
Fermi level and well isolated from the rest of the spectrum.
A typical example of the electronic structure of YVO$_3$, obtained in the
local-density approximation (LDA), is shown in Fig.~\ref{fig.YVO3LDA}.
In this case, the active states are
twelve $t_{2g}$ bands of predominantly V-character, which are separated by finite energy
windows from other states, both from below and from above.
Such a division of the entire electronic structure into the ``active'' and
``inactive'' (or low- and high-energy)
states opens a formal way for combining the first-principles
calculations, based on the density-functional theory (DFT),
with the many-body treatment of some effective model, formulated rigorously in the restricted Hilbert
space of ``active'' states.
This is the main idea of realistic
modeling of complex oxide materials.
The purpose of this project is twofold:

(1)
To incorporate the physics of Coulomb correlations,
which is greatly oversimplified
in conventional LDA;

(2)
To provide a transparent physical picture for electronic and magnetic properties of complex compounds.
In this sense, the realistic modeling can be regarded as supplementary approach to more conventional
first-principles electronic structure calculations, which are currently on the rise.

  Thus, the first step of the project is construction of the low-energy model
(typically, the multiorbital Hubbard model) for the states near the Fermi level:
\begin{equation}
\hat{\cal{H}}  =  \sum_{{\bf RR}'} \sum_{\alpha \beta}
t_{{\bf RR}'}^{\alpha \beta}\hat{c}^\dagger_{{\bf R}\alpha}
\hat{c}^{\phantom{\dagger}}_{{\bf R}'\beta} +
  \frac{1}{2}
\sum_{\bf R}  \sum_{\alpha \beta \gamma \delta} U_{\alpha \beta
\gamma \delta} \hat{c}^\dagger_{{\bf R}\alpha} \hat{c}^\dagger_{{\bf R}\gamma}
\hat{c}^{\phantom{\dagger}}_{{\bf R}\beta}
\hat{c}^{\phantom{\dagger}}_{{\bf R}\delta},
\label{eqn.ManyBodyH}
\end{equation}
which would include the effect of other (``inactive'') states
in the definition of
the model parameters of the Hamiltonian (\ref{eqn.ManyBodyH}). All these parameters
are derived totally from the ``first-principles'' on the basis of DFT.
We would like to emphasize that,
although the derivations are
inevitably based on some approximations,
we do not use any adjustable parameters apart from these
approximations.
The procedure of constructing the
model Hamiltonian was described in details in the review article \cite{review2008}. Briefly,
in order to derive
the one-electron part ($t_{{\bf RR}'}^{\alpha \beta}$), we use the generalized
downfolding method. For the isolated low-energy bands, this procedure is exact, as it is
clearly seen in Fig.~\ref{fig.YVO3LDA} from the comparison of the
original band structure, obtained in the linear muffin-tin orbital method (LMTO) \cite{LMTO}, and the one after
the downfolding. The parameters of screened Coulomb interactions
($U_{\alpha \beta \gamma \delta}$) are typically obtained by combining the
constrained DFT technique \cite{Dederichs} with the random-phase approximation \cite{Ferdi04}.
The latter is very efficient for treating the screening of correlated electrons by
themselves \cite{PRL05}, which can dramatically reduce the effective Coulomb interactions in the
low-energy bands. For example, the bare Coulomb repulsion between $3d$ electrons is
about
20-25 eV. However, for the low-energy bands in solids this value is
typically reduced till 2-4 eV~\cite{review2008}.

  Once the model is constructed, it can be solved by using various many-body techniques.
In the present work, we typically start with the mean-field Hartree-Fock (HF) approximation, and
take into account the correlation interactions by considering the
perturbation theory expansion near the HF ground state.
This procedure may be justified, if the degeneracy of the ground state is lifted
by the crystal distortions \cite{review2008}.

  Below, we present examples of realistic modeling for two types of
transition-metal oxides.

\section{Spin-Orbital-Lattice Coupling in vanadates $R$VO$_3$}
\label{sec.AVO3}

\begin{figure}[!h]
\centering
\includegraphics[width=1.5in]{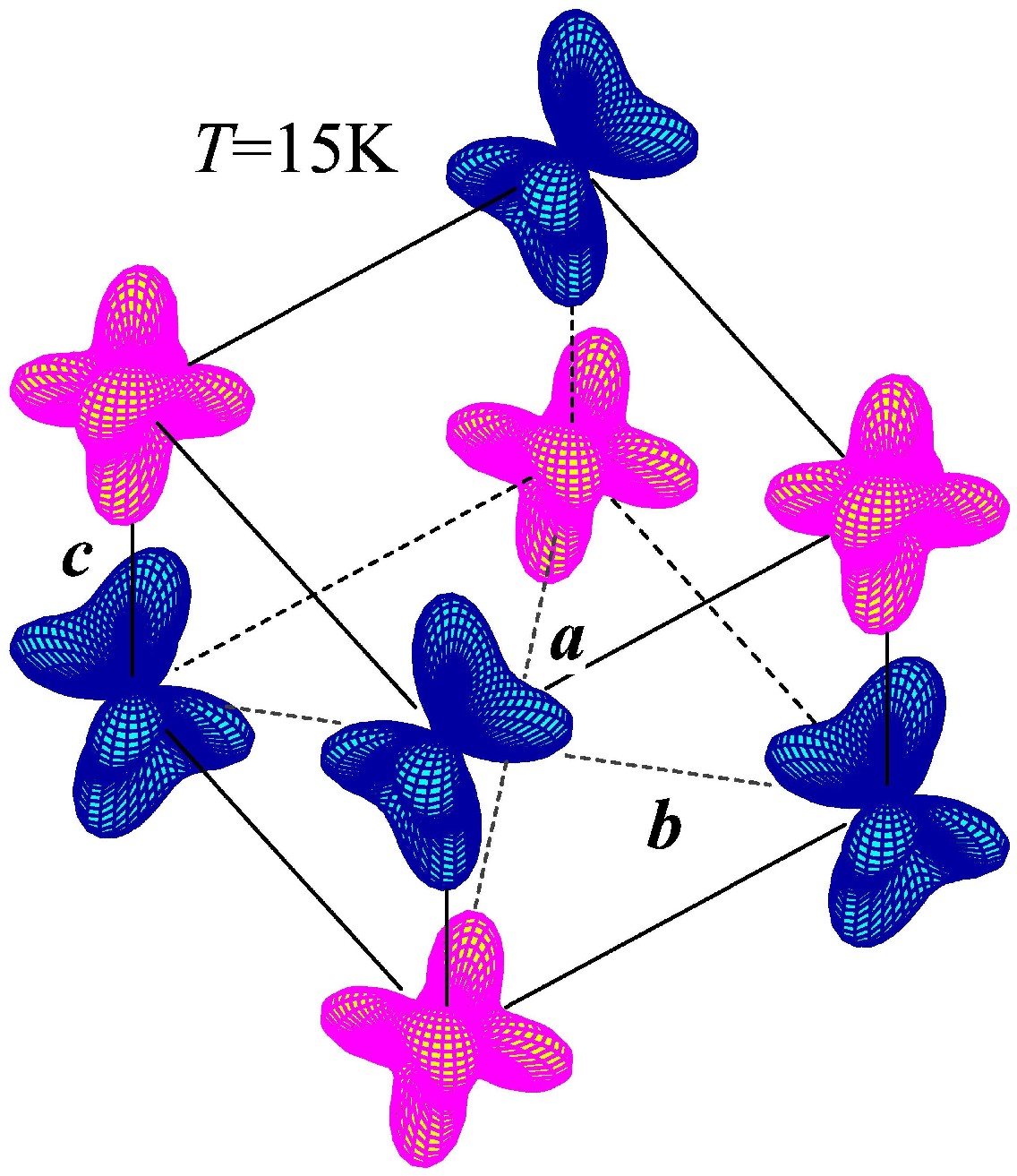} \includegraphics[width=1.5in]{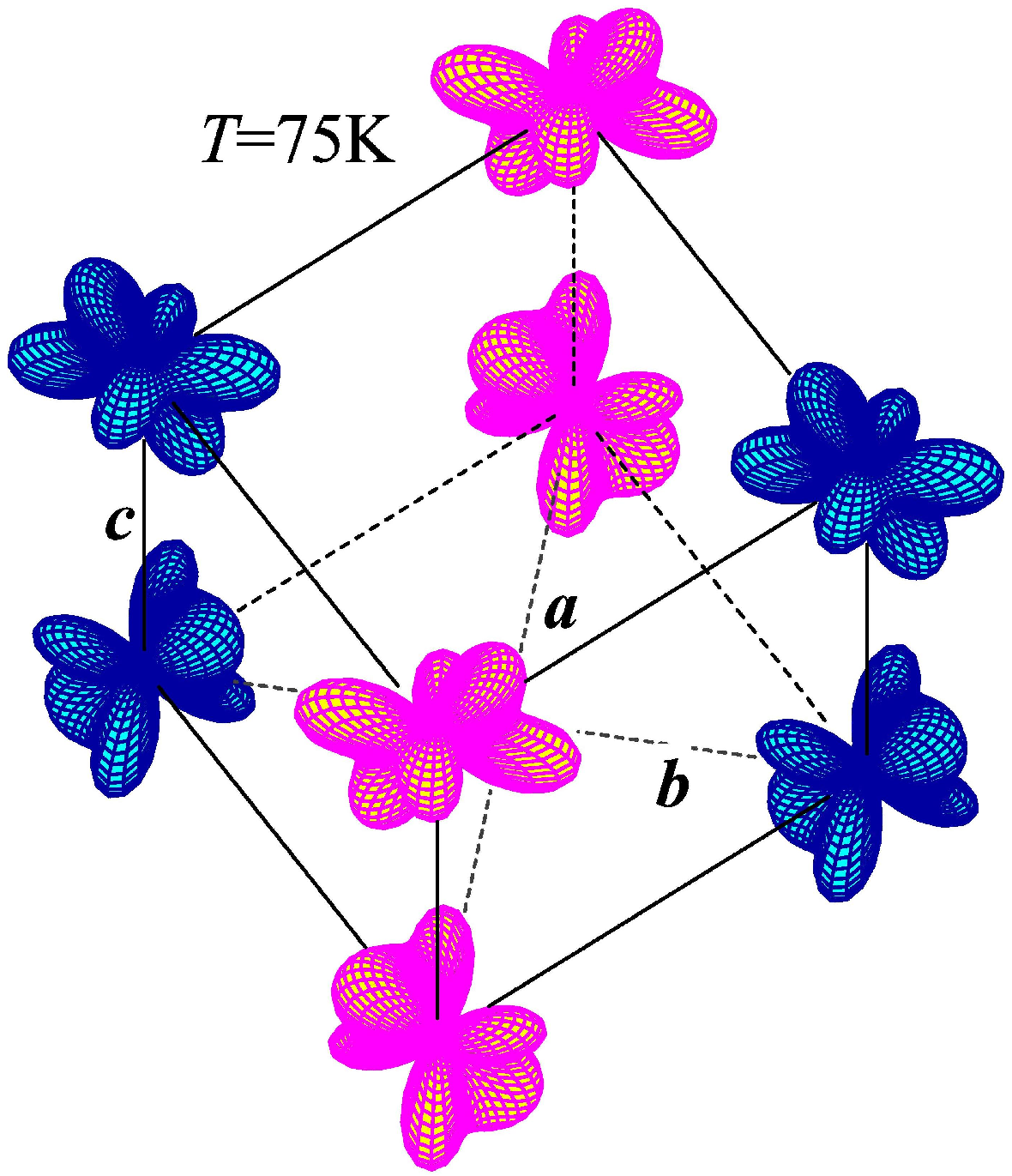}
\caption{Distribution of the charge densities associated with the occupied
$t_{2g}$ orbitals (the orbital ordering) realized in the orthorhombic (left)
and monoclinic (right) phase of YbVO$_3$ in the HF approximation.
Different magnetic sublattices associated with two opposite
directions of spins in the AFM structure are shown by different colors.}
\label{fig.YbVO3OO}
\end{figure}

\begin{figure}[!h]
\centering
\includegraphics[width=3.0in]{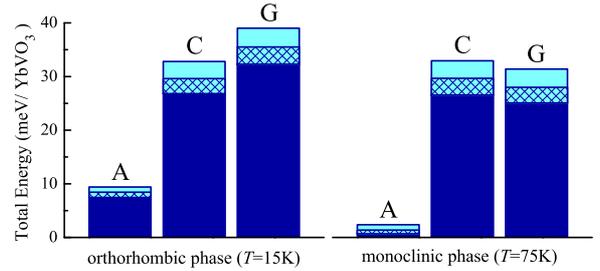}
\caption{Stabilization energies of the main AFM states in YbVO$_3$
relative to the ferromagnetic state as obtained in the HF approximation (dark blue area)
and after taking into account the correlation interactions in the second order of
perturbation theory (light blue area) and in the t-matrix theory (hatched area).}
\label{fig.YbVO3E}
\end{figure}

  The vanadates $R$VO$_3$ (where $R$ is the three-valent, typically rare-earth, element)
have attracted a considerable experimental and theoretical attention.
All these compounds crystallize in the distorted perovskite structure.
For example, considered in the present work YbVO$_3$ can have orthorhombic and
monoclinic modifications, realized at $T$$=$ 15 and 75 K, respectively. However, a relatively
small change of the lattice parameters may cause a dramatic reconstruction
of electronic, and associated to it, magnetic structure. For example, at $T$$=$ 15 and 75 K
YbVO$_3$ forms the so-called G- and C-type antiferromagnetic (AFM) structure,
respectively (Fig.~\ref{fig.YbVO3OO}).
The phenomenon called ``spin-orbital-lattice coupling'' and is typically regarded as the
test for various theories of the electronic structure. First attempts of realistic
modeling of $R$VO$_3$ are summarized in~\cite{review2008}. After that we have
performed a systematic study for the entire series $R$VO$_3$
($R$$=$ La, Ce, Pr, Nd, Sm, Gd, Tb, Ho, Yb, Lu, and Y), using the experimental crystal structure
available in the literature~\cite{RVO3_structure1,RVO3_structure2}. For all considered compounds,
the effective Hubbard model was constructed for $t_{2g}$ bands (Fig.~\ref{fig.YVO3LDA}) in the basis of three
Wannier orbitals per each V-site.
The details will be
published elsewhere. Here we present an example of the ``canonical behavior'' realized in
YbVO$_3$, where the crystal distortion quenches the orbital structure in some
particular configuration (Fig.~\ref{fig.YbVO3OO}),
which uniquely defines the type of the magnetic ground state,
basically via Goodenough-Kanamori rules. In this case, once the degeneracy of
$t_{2g}$-levels is lifted
by the crystal distortion, the correct type of the magnetic ground state can be successfully reproduced
already at the level of HF approximation. For example, the stabilization energy is clearly the largest
for the experimentally observed G- and C-type AFM states when YbVO$_3$ crystallizes in the
orthorhombic and monoclinic structure, respectively (Fig.~\ref{fig.YbVO3E}).
Moreover, the correct AFM ground state is additionally stabilized by correlation interactions,
which are taken into account via perturbation-theory expansion near the HF solutions. We have considered
two such technique for the total energy~\cite{JETP07}: one is the regular second-order perturbation theory and
the other one is the t-matrix theory.
Both of them provide very consistent explanation for YbVO$_3$, although the
correlation energies obtained in the t-matrix theory for the AFM states are
systematically smaller due to the higher-order correlation effects, which are
included to the t-matrix, but not to the second-order perturbation theory.

\section{Inversion Symmetry Breaking in Manganites $R$MnO$_3$}
\label{sec.multiferroicity}

  The multiferroic manganites, such as BiMnO$_3$ and TbMnO$_3$,
are currently under very intensive investigation. The multiferroicity means that
the magnetic order in certain system without the inversion symmetry coexists with some finite ferroelectric polarization.
The coupling of these two order parameters
provides a unique opportunity to control the magnetic properties by applying the
electric field and vice versa. From this point of view, the most interesting is the situation,
where the inversion symmetry is broken by
the magnetic degrees of freedom.
In order to study the microscopic origin of the inversion symmetry breaking in manganites, we have
constructed the effective model in the basis of
three $t_{2g}$ and two $e_g$ orbitals at each Mn-site \cite{JPSJ,JETP09}.
The main results can be summarized as follows.

\begin{figure}[!h]
\centering
\includegraphics[width=3.0in]{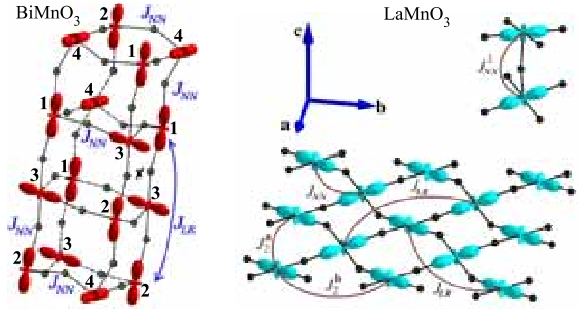}
\caption{Typical distribution of the charge densities associated with the occupied
$e_g$ orbitals (the orbital ordering) realized in the monoclinic (BiMnO$_3$, left)
and orthorhombic (LaMnO$_3$, right) manganites
with the notations of main magnetic interactions.}
\label{fig.OrbitalOrderManganites}
\end{figure}

\begin{figure}[!h]
\centering
\includegraphics[width=3.0in]{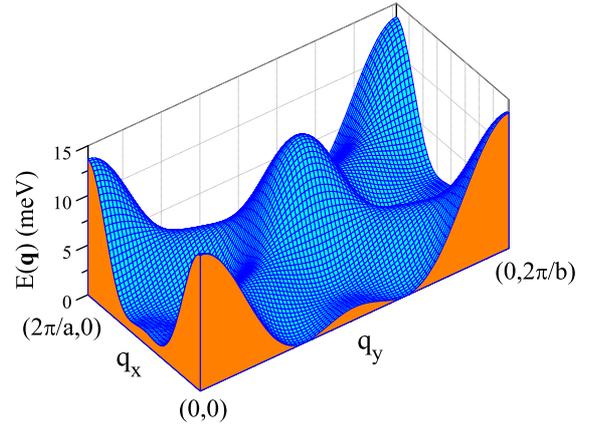}
\caption{Dependence of the total energy
(measured in meV per one formula unit)
on the homogeneous spin-spiral
vector ${\bf q}$ obtained in the Hartree-Fock approximation for TbMnO$_3$ without
the relativistic spin-orbit coupling.}
\label{fig.TbNbO3Eq}
\end{figure}

  The mechanism of the inversion symmetry breaking is related to the
behavior of interatomic magnetic interactions,
which depends on the distribution of occupied
$e_g$ orbitals (Fig.~\ref{fig.OrbitalOrderManganites}).
Although
the orbital ordering differs dramatically in monoclinic and
orthorhombic manganites, the basic idea is rather generic: the nearest-neighbor interactions $J_{NN}$
always compete with some long-range AFM interactions $J_{LR}$. The existence of $J_{LR}$ is
related to the fact that, due to the screening, the on-site Coulomb repulsion $U$
is not particularly large (about 2.2 eV in manganites). Therefore, other mechanisms
of magnetic interactions, besides conventional
superexchange (being of the order of $1/U$), can be also operative. These mechanisms,
which are of the higher orders than $1/U$, together with the form of the orbital ordering,
naturally explain the large values of $J_{LR}$
in some particular bonds \cite{JPSJ}. When $J_{LR}$ becomes larger than $J_{NN}$,
they can lead to the
formation of complex magnetic structures with the broken inversion symmetry.
In monoclinic BiMnO$_3$, this is
the collinear AFM structure, where each of the two sublattices (1,2) and (3,4) are
coupled antiferromagnetically
(see Fig.~\ref{fig.OrbitalOrderManganites} for the notations of Mn-sites)~\cite{JETP09}.
In TbMnO$_3$, the ground state is the incommensurate spin-spiral,
where the directions of spins in the ${\bf ab}$-planes vary as
${\bf e}_{\bf R}$$=$$(\cos {\bf qR},\sin {\bf qR}, 0)$, and the spin moments between the
planes are coupled antiferromagnetically. Such a magnetic structure can be easily calculated
by using the generalized Bloch theorem, which combines the lattice translations with the
spin rotations~\cite{Sandratskii}. The total energy minimum
corresponds to the propagation along the orthorhombic ${\bf b}$-axis, in agreement with the
experiment. However, the obtained value $q_b$$=$$0.675$ (in units $\pi/b$, Fig.~\ref{fig.TbNbO3Eq}),
is substantially larger than
experimental $q_b$$=$ $0.25$~\cite{Arima} and $0.28$~\cite{Kimura}, reported for
magnetic structures with the moments lying in the ${\bf ab}$- and ${\bf bc}$-plane,
respectively.

\begin{figure}[!h]
\centering
\includegraphics[width=3.0in]{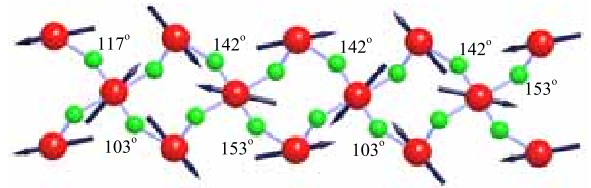}
\caption{Distribution of magnetic moments in the ${\bf ab}$-plane
obtained in the HF approximation with SOI
for TbMnO$_3$. The angles formed by Mn-moments in the bonds Mn-O-Mn are specified by numbers.}
\label{fig.TbNbO3SO}
\end{figure}

  In order to resolve this discrepancy,
we consider the relativistic spin-orbit interaction (SOI). In BiMnO$_3$,
the SOI is responsible for the spin canting away from the collinear AFM structure
and formation of the net ferromagnetic moment. Thus, the ferroelectric response in BiMnO$_3$, caused by the
hidden AFM order, coexists with the ferromagnetism and can be controlled by the magnetic field~\cite{JETP09}.
The magnetic interactions of the relativistic origin play an important role also in TbMnO$_3$~\cite{Mochizuki}.
In the search for the true magnetic ground state of TbMnO$_3$, we have investigated several magnetic structures with
different periodicity. The structure corresponding to the lowest energy is shown in Fig.~\ref{fig.TbNbO3SO}.
The periodicity of this structure
is described by $q_b$$=$$0.25$, in agreement with the experiment~\cite{Arima}. Nevertheless,
it is no longer the
uniform spin-spiral.
It would be interesting to check our finding experimentally.

\section{Conclusions}
\label{sec.conclusions}

The realistic modeling combines the accuracy and predictable power of first-principles electronic structure calculations with the flexibility and insights of the model analysis. The first applications of this method are very encouraging. We hope that these ideas will continue to develop to become a powerful tool for theoretical analysis of complex oxide materials and other strongly correlated systems.

\section*{Acknowledgment}
The work is partly supported by Grant-in-Aid for Scientific
Research (C) No. 20540337 and
in Priority Area ``Anomalous Quantum Materials''
from the
Ministry of Education, Culture, Sport, Science and Technology of
Japan.

%% The Appendices part is started with the command \appendix;
%% appendix sections are then done as normal sections
%% \appendix

%% \section{}
%% \label{}

%% References
%%
%% Following citation commands can be used in the body text:
%% Usage of \cite is as follows:
%%   \cite{key}         ==>>  [#]
%%   \cite[chap. 2]{key} ==>> [#, chap. 2]
%%

%% References with bibTeX database:

\bibliographystyle{elsarticle-num}
%%\bibliography{<your-bib-database>}

%% Authors are advised to submit their bibtex database files. They are
%% requested to list a bibtex style file in the manuscript if they do
%% not want to use elsarticle-num.bst.

%% References without bibTeX database:

\end{document}